\documentclass[showpacs]{revtex4-2}
\usepackage{graphicx} 
\usepackage{subfigure}
\usepackage{multirow}
\usepackage{fancyhdr}
\usepackage{longtable}
\usepackage{parskip}
\usepackage[T1]{fontenc}
\usepackage{dcolumn} 
\usepackage{bm}       
\usepackage{amsfonts}  
\usepackage{amsmath}   
\usepackage{amssymb}   
\usepackage[]{hyperref}

\newcommand{\omegavec}{\boldsymbol{\omega}}
\newcommand{\thetavec}{\boldsymbol{\theta}}

\begin{document}

\title{Synchronization of  frustrated phase oscillators in the  small-world networks}
\author{Esmaeil Mahdavi}

\author{Mina Zarei}
\affiliation{\it Department of Physics, Institute for Advanced Studies in Basic Sciences (IASBS), Zanjan 45137-66731, Iran}

\author{Farhad Shahbazi} 
\affiliation{\it Department of Physics, Isfahan University of Technology, Isfahan 84156-83111, Iran}

\begin{abstract}  

We numerically study the synchronization of an identical population of Kuramoto-Sakaguchi phase oscillators in Watts-Strogatz networks. 
We find that, unlike random networks, phase-shift could enhance the synchronization in small-world networks. We also observe abrupt phase transition with hysteresis at some values of phase shifts in small-world networks, signs of an explosive phase transition. Moreover, we report the emergence of {\em Chimera} states at some values of phase-shift close to the transition points, which consist of spatially coexisting synchronized and desynchronized domains. 
\end{abstract}

\maketitle

\section{Introduction}

The emergence of synchronization in a network of coupled nonlinear oscillators plays a crucial role in many real biological, chemical, mechanical, and ecological systems~\cite{mirollo1990synchronization,garcia2004modeling,tyson1973some,nijmeijer2003synchronization,blasius1999complex}. A limit-cycle oscillator in these systems is often modeled as nonlinear multi-dimensional differential equations. However, through phase reduction theory, one can systematically simplify these multi-dimensional differential equations to a one-dimensional phase equation that approximately describes the oscillator dynamics~\cite{nakao2016phase}. One of the most studied models for phase synchronization is the Kuramoto model, where the synchronization mechanism is due to a nonlinear coupling associated with the Sinus of the phase difference between oscillators~\cite{kuramoto1975self,kuramoto2012chemical}.

The interplay between the structure and dynamics in complex networks has been attracting much interest within the scientific community for several decades~\cite{arenas2008synchronization,rodrigues2016kuramoto}. Several efforts have been made to find the optimal structure for synchronization of the coupled identical and non-identical oscillators. Most of these studies used linear reformulation of the synchronization model, particularly the Kuramoto model, to obtain their results~\cite{skardal2014optimal}. However, linearization gives rise to a significant error for the Kuramoto model  on modular and small-world (SW ) networks~\cite{ghorban2021linearization}.

Small-world properties make a network very efficient in communication and are the characteristics of many real-world systems. The SW networks have two primary features: a short average shortest path length and a high clustering coefficient. A well-known random rewiring procedure was introduced by Watts and Strogatz that produces random graphs with SW properties by interpolating between a regular ring lattice and a random network~\cite{watts1998collective,watts2001small}. 

Previous studies have revealed that the Kuramoto model in SW networks displays extremely rich dynamical behaviors~\cite{barahona2002synchronization,esfahanii2012noise,nikfard2021enhancement,ameli2021time2}. For instance, it has turned out that an SW network of identical phase oscillators interacting by the Kuramoto coupling has various attractors characterized by different integer winding numbers. Starting from different initial phase distributions, the stationary state exhibits some phase-locked patterns in the form of isolated defects or quasi-periodic states~\cite{esfahanii2012noise}. Moreover, it has been shown that noise~\cite{esfahanii2012noise,nikfard2021enhancement} and time delay~\cite{ameli2021time2} can enhance synchronization in the SW networks by washing out those phase-locked patterns.
 
To describe the realistic oscillatory systems, modified phase coupling models seem to be more relevant than the original ones. To this end, the introduction of a frustration parameter  to the  coupling has attracted increasing attention in recent years. Previous researches have reported that frustration, in general, can enrich and diversify the dynamics of the phase models such as Heisenberg XY model~\cite{yokoi1988ground}, Frenkel-Kontorova model~\cite{zheng1998resonant} and Josephson junction arrays~\cite{watanabe1996dynamics}.

The Sakaguchi-Kuramoto model~\cite{sakaguchi1988mutual,sakaguchi1986soluble} is a modification of the well-known Kuramoto model, in which a frustration parameter in the form of a phase shift is added to the coupling of each phase oscillator's pair. 
Quantitative and qualitative studies have shown that including even a small amount of frustration makes the analysis of the model more difficult than the original Kuramoto model. For example, the existence of frustration destroys a gradient flow structure of the original Kuramoto equation~\cite{ha2018remarks}.

A constant phase shift can be considered as an approximation for modeling time-delayed couplings when the delay is small~\cite{crook1997role}. A method based on the linearization of the frustrated Kuramoto model is applied to evaluate the global synchrony in coupled oscillators in the presence of constant frustration. The results show that even in the limit of infinite coupling, perfect phase synchronization is unattainable in the steady state~\cite{skardal2015erosion}. In addition, employing Ott-Antenson method, unusual types of synchronization transitions have been reported for the globally coupled oscillators with constant frustration~\cite{omel2012nonuniversal}. Another study has shown that one can enhance synchronization by tuning frustrations using local connectivity and the intrinsic frequency of each oscillator~\cite{brede2016frustration}.  
Moreover, the Kuramoto-Sakaguchi model has been used as a basis for studying chimera states, where identical symmetrically coupled oscillators display a remarkable spatio-temporal pattern in which regions of coherence and incoherence coexist~\cite{scholl2016synchronization,abrams2004chimera,abrams2008solvable,laing2009chimera,yeldesbay2014chimeralike,martens2016chimera}.

In this work, we investigate the dynamics of the frustrated Kuramoto model in Watts-Strogatz networks of identical phase oscillators. As we mentioned before, the linearization error is too large for SW topologies. So we can not necessarily extend the results found by linearization for the SW networks. We show that small values of the frustration parameter could increase the synchronization by weakening the defects. Furthermore, we found chimera states for the frustration parameters for which the oscillator's frequencies start to spread with skewed distributions. 
This paper is organized as follows: 
in section~\ref{model} we introduce the model and method of characterizing the global and local synchronization in the system. 
We discuss the numerical results in SW and random networks in section~\ref{results}, and section~\ref{conclusion} is devoted to the conclusion.

\section{Model and method}
\label{model}
In this section, we briefly describe the networks and  models that we used in the paper.

\subsection{Network Model}
In this paper, we study the phase synchronization of SW and random networks in the presence of frustration. Therefore, the Watts-Strogatz (WS) algorithm~\cite{watts1998collective,watts2001small} has been used to construct the networks. According to this algorithm, starting from a regular ring lattice of $N$ nodes, each connected to its $K$ nearest neighbors, the links are rewired with probability $p$. In this way, $p=0$ describes the regular network, while $p=1$ leads to the random networks. Furthermore, for the small rewiring probability, $0.005 \lesssim p \lesssim 0.05$, SW networks appear. These small-world networks have two main desired features: short average shortest path lengths and high clustering coefficients. Throughout this paper, we constructed our networks with $N=1000$, $\langle K\rangle=10$, and for the SW networks, we use the rewiring probability $p=0.03$.
 
\subsection{Synchronization Model}
\label{synchronizationmodel}
The Sakaguchi-Kuramoto model is a modification of the well-known Kuramoto model, in which a frustration factor is added to a network of phase-coupled oscillators. The model is defined as,
		\begin{eqnarray}
			\frac{d\theta_{i}}{dt} & = &\omega_{i}+\lambda\sum_{j=1}^N a_{ij} \sin(\theta_{j}(t)-\theta_{i}(t)-\alpha),
		\label{Eq:sakaguchikuramoto} \end{eqnarray}	
		here  $ \theta_{i} $ and $ \omega_{i} $, denote the phase and intrinsic frequency of the $i$th oscillator.  $N$, $ \lambda $ and $\alpha$ represent the number of   oscillators, coupling constant and the frustration parameter, respectively.  The elements of the adjacency matrix are indicated by $  a_{ij} $'s, where $a_{ij}=1$ indicates direct link between elements  and $a_{ij}=0$ indicates indirect relationship.
		
Linearizing  the Sakaguchi-Kuramoto model around the synchronized state gives:
\begin{equation}
\label{linear}
\frac{d\theta_i}{dt}=\omega_i-\lambda d_{i}\sin\alpha+\lambda\sum_{j=1}^{N}a_{ij}(\theta_{j}-\theta_{i})\cos\alpha,
\end{equation}
in which $d_i$ is the degree on node $i$ and   $i=1,\ldots,N$. \\
Setting $\omega'_i=\omega_i-\lambda d_i\sin\alpha$ and  
$\lambda'=\lambda\cos\alpha$, Eq.\eqref{linear} can be rewritten as:
\begin{align}\label{frustLinear}
	\dfrac{d \thetavec}{dt}=\omegavec'- L'\thetavec,
\end{align}
where
$\omegavec'=[\omega'_1\ \cdots \ \omega'_N]^T$, and $L'$ is the Laplacian matrix for the weighted adjacency matrix $A'=\lambda' A$. Therefore, the linearized frustrated Kuramoto model resembles the linearized ordinary Kuramoto model with frustration-dependent coupling and intrinsic frequencies. 
A recent study shows that the linearization error would be very large for the SW networks~\cite{ghorban2021linearization}. 
Using the linearization method,  Ghorban et al.~\cite{ghorban2021linearization} found a lower bound for the order parameter of identical Kuramoto-Sakaguchi oscillators, which is decreased by frustration.

The level of global synchrony in a network is quantified by the order parameter ($r$), which is determined by the following equation,
		\begin{eqnarray}
			r (t)& = & \frac{1}{N} \mid \sum_{j=1}^N e^{i\theta_{j}(t)}\mid,
                    \label{Eq:r}
		\end{eqnarray}
where $0\leq r \leq 1$, in which $r=1$ shows the fully synchronized state, while $r=0$ indicates the incoherent state. The time average of $r$ after achieving the steady state is represented by $r_{\infty}$.
To get the structure of phase configurations in the network, correlation matrix~($D$) is defined as follows
		\begin{eqnarray}
			D_{ij} & = &\lim_{	\Delta t \to \infty} \frac{1}{\Delta t} \int_{t_{s}}^{t_{s}+\Delta t} \cos(\theta_{i}(t)-\theta_{j}(t)) \;\mathrm{d}t
                     \label{Eq:dij}
		\end{eqnarray}
where, $t_s$ is the time to reach steady state~\cite{gomez2007paths}.  The elements of the correlation matrix are in the range of $[-1,1]$, in which  $+1$ and  $-1$ indicate in-phase and anti-phase relationships between any two nodes $i$ and $j$. {Throughout this paper, we used $2.5\times 10^5$ time steps for the stationary time $t_s$ and $1.5\times 10^5$  time steps for the averaging window $\Delta t$.}

The local order parameter, which measures the degree of synchrony for each node and its $2m$ nearest neighbors along the initial ring-shaped network, can be calculated from the following equation
		\begin{eqnarray}
			R_{i}(t) & = & \frac{1}{2m} \mid \sum_{j=i-m}^{i+m} e^{i\theta_{j}(t)}\mid
			  \label{Eq:Rij}
		\end{eqnarray}
where $R_i$ lies in the interval $[0,1]$, in which $0$ and $1$ indicate whether the $i$th node is synchronized with its neighbors or not, respectively..

Finally, the angular frequency of the $i$th oscillator in the the rotating frame is calculated by  $\tilde{\omega}_{i}=\dot\theta_{i}-\langle \omega_{i}\rangle$. Therefore, the average collective angular frequency over a long-time period in the rotating frame is defined as,

\begin{equation}
\Omega= \frac{1}{N}\sum_{i=1}^{N}\lim_{\Delta t \to \infty}\frac{1}{\Delta t}\int^{\Delta t+t_s}_{t_s}\tilde{\omega}_i.
\end{equation}
where, $t_s$ is the time to reach steady state.

Throughout this paper, the fourth-order Runge-Kutta integration method is used for numerical simulation with a time 
step size 0.02. Moreover, it is assumed that all oscillators have equal intrinsic frequencies,$\omega_i=\omega_0$ 
which can be set to zero by moving to the rotating frame. Rescaling the time as $t'=\lambda t$ makes the steady states independent of the value of the coupling constant, so allowing us to fix the coupling constant 
$\lambda$ to unity.

\section{Results and discussion}
\label{results}

\begin{figure}[]
\centering
 {\includegraphics[width=0.5\columnwidth]{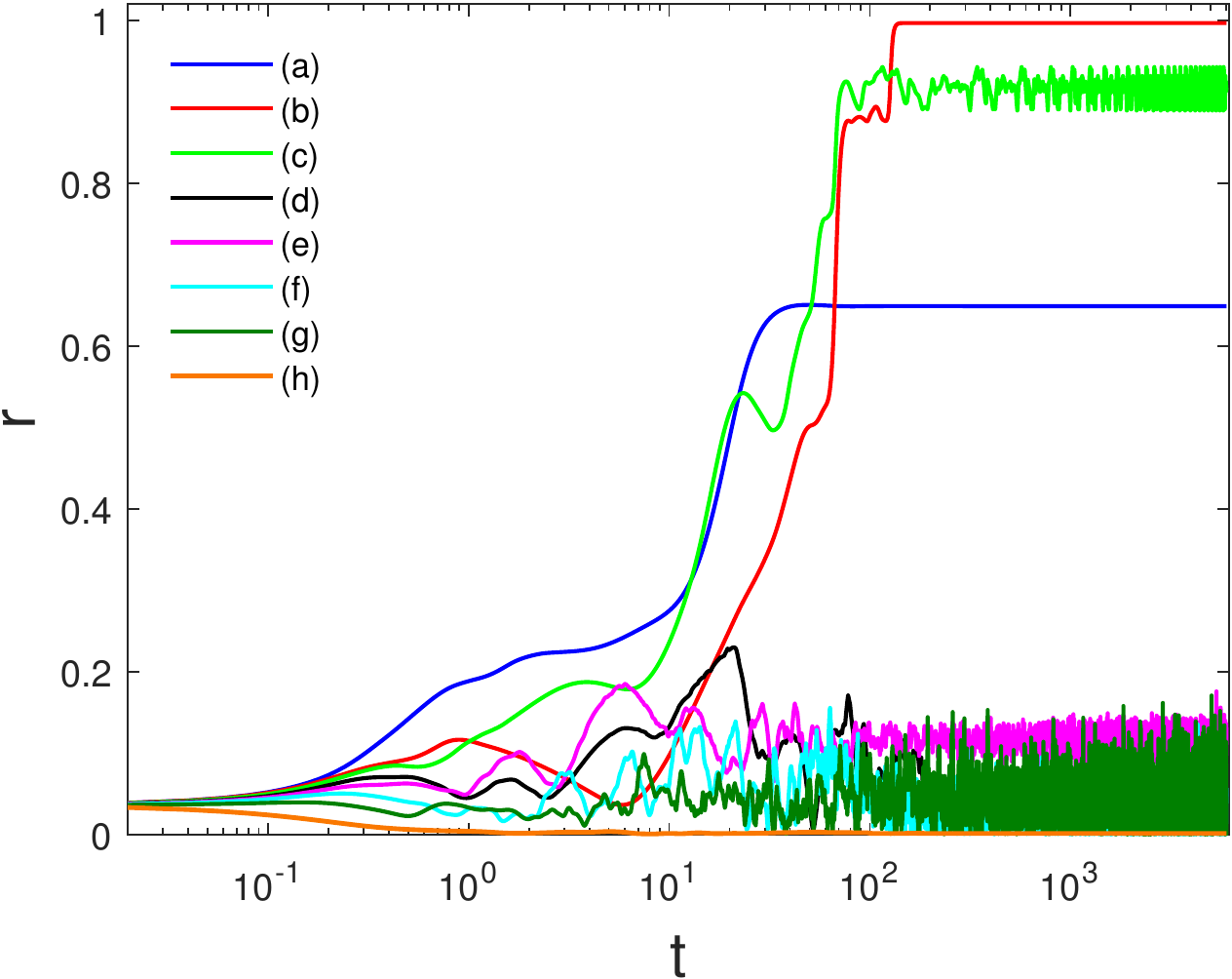}}
\caption{The time evolution (in logarithmic scale) of the order parameter for a SW network with various frustration parameters. Different curves correspond to  (a) $\alpha=0$, (b) $\alpha=0.1\pi$, (c) $\alpha=0.2\pi$, (d) $\alpha=0.3\pi$, (e) $\alpha=0.35\pi$, (f) $\alpha=0.4\pi$, (g) $\alpha=0.5\pi$, and (h) $\alpha=\pi$.  Except frustration, similar parameters and initial phase values have been assigned for all the curves. The network parameters are $N=1000$, $ \langle k \rangle = 10 $, $p=0.03$. }
\label{r-t-WS}
\end{figure}
\begin{figure}[]
\centering
 {\includegraphics[width=0.5\columnwidth]{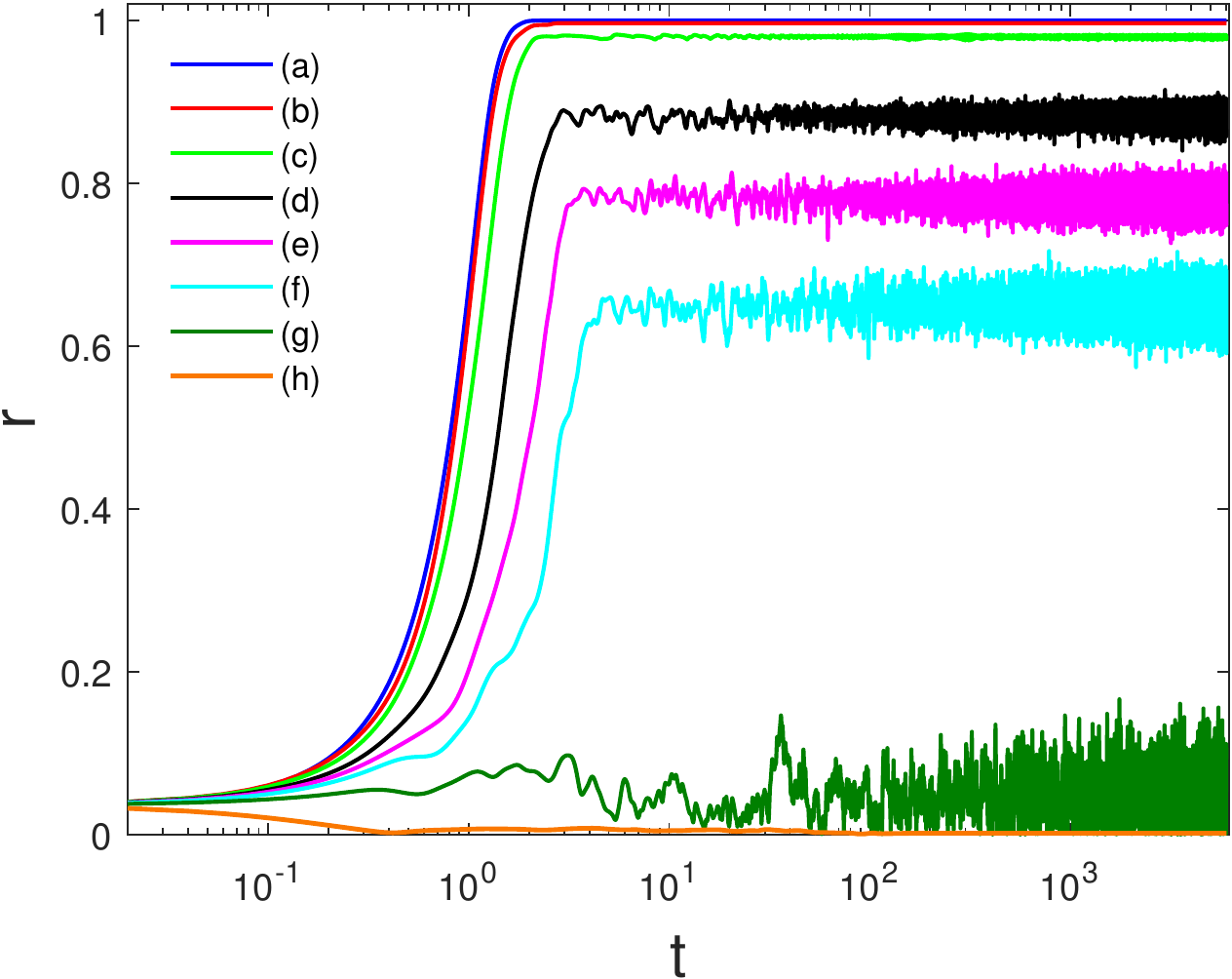}}
\caption{The time evolution (in logarithmic scale) of the order parameter for a random network with various frustration parameters. Different curves correspond to  (a) $\alpha=0$, (b) $\alpha=0.1\pi$, (c) $\alpha=0.2\pi$, (d) $\alpha=0.3\pi$, (e) $\alpha=0.35\pi$, (f) $\alpha=0.4\pi$, (g) $\alpha=0.5\pi$, and (h) $\alpha=\pi$.  Other parameters and initial phase values are those used in Fig.~\eqref{r-t-WS}.}
\label{r-t-R}
\end{figure}
\begin{figure}[]
\centering
{\includegraphics[width=\textwidth]{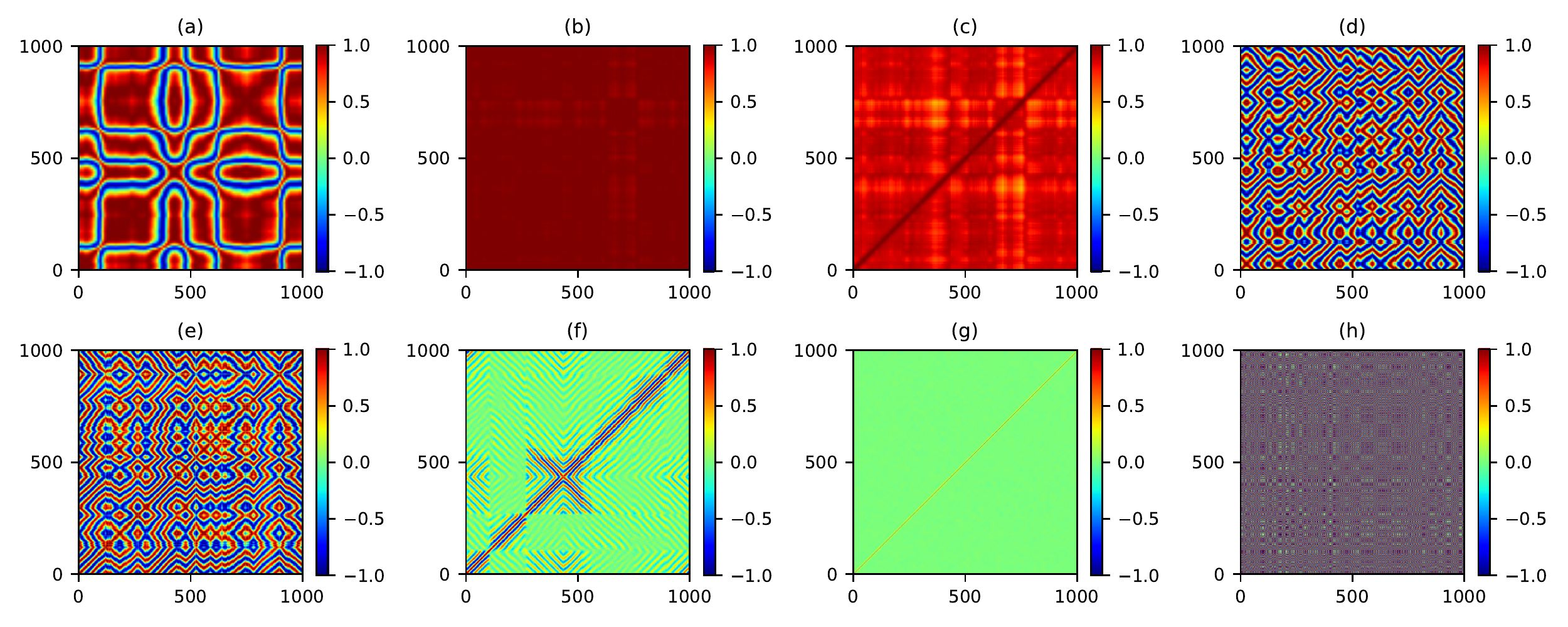}}
\caption{Density plots of the correlation matrix elements in a SW network for various frustration parameters. Different panels correspond to  (a) $\alpha=0$, (b) $\alpha=0.1\pi$, (c) $\alpha=0.2\pi$, (d) $\alpha=0.3\pi$, (e) $\alpha=0.35\pi$, (f) $\alpha=0.4\pi$, (g) $\alpha=0.5\pi$, and (h) $\alpha=\pi$.  The axes show label of oscillators, and the colorbars display the values of correlations as defined in Eq.~\eqref{Eq:dij}. Other parameters and initial phase values are those used in Fig.~\ref{r-t-WS}.}
\label{CSW}
\end{figure}
\begin{figure}[]
\centering
 {\includegraphics[width=\textwidth]{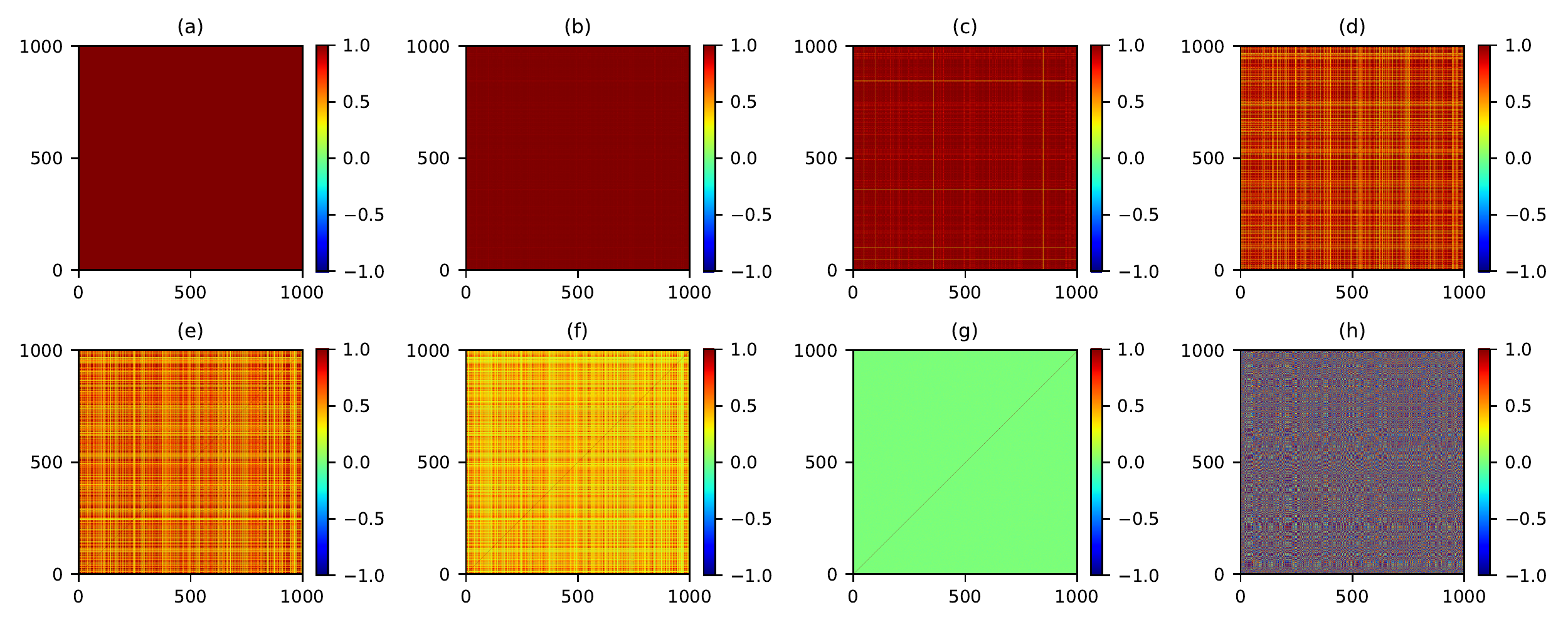}}
\caption{Density plots of the correlation matrix elements in a random network for various frustration parameters. Different panels correspond to  (a) $\alpha=0$, (b) $\alpha=0.1\pi$, (c) $\alpha=0.2\pi$, (d) $\alpha=0.3\pi$, (e) $\alpha=0.35\pi$, (f) $\alpha=0.4\pi$, (g) $\alpha=0.5\pi$, and (h) $\alpha=\pi$.  The axes show label of oscillators, and the colorbars display the values of correlations as defined in Eq.~\eqref{Eq:dij}. Other parameters and initial phase values are those used in Fig.~\ref{r-t-R}.}
\label{CR}
\end{figure}

\begin{figure}[]
\centering
 {\includegraphics[width=\textwidth]{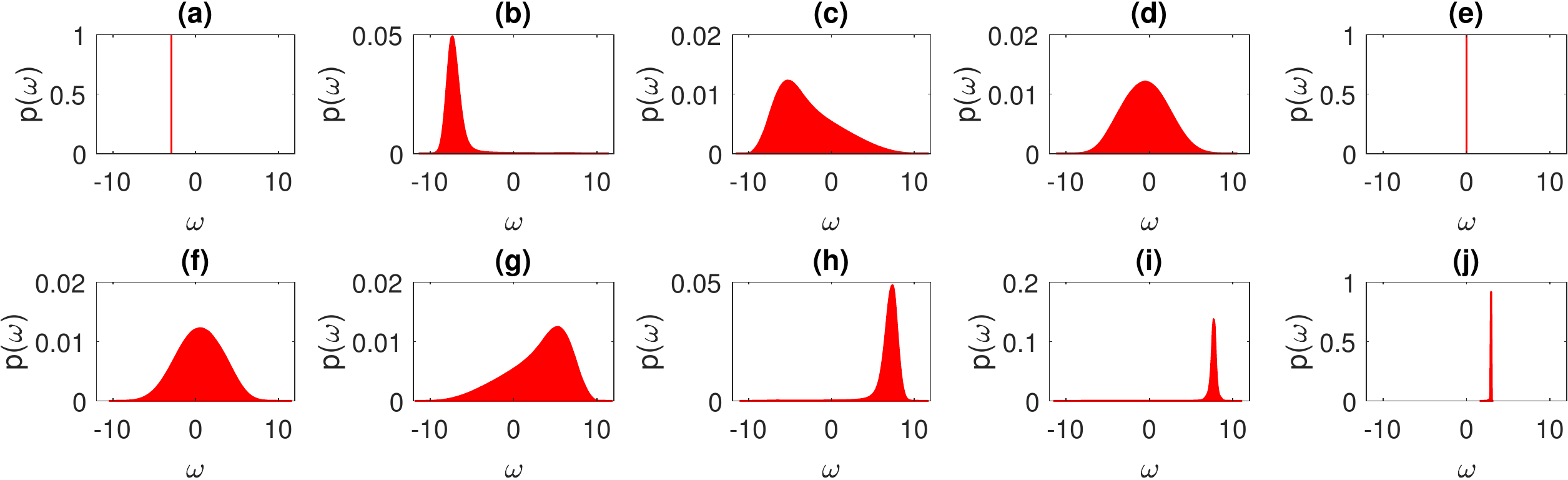}}
\caption{The probability density function of the stationary frequencies of the oscillators in a SW network with different frustrations.  Different panels correspond to  (a) $\alpha=0.1\pi$, (b) $\alpha=0.4\pi$, (c) $\alpha=0.45\pi$, (d) $\alpha=0.5\pi$, (e) $\alpha=\pi$, (f) $\alpha=1.5\pi$, (g) $\alpha=1.55\pi$, (h) $\alpha=1.6\pi$, (i) $\alpha=1.64\pi$, and (j) $\alpha=1.9\pi$.  }
\label{DF-SW}
\end{figure}
\begin{figure}[]
\centering
 {\includegraphics[width=\textwidth]{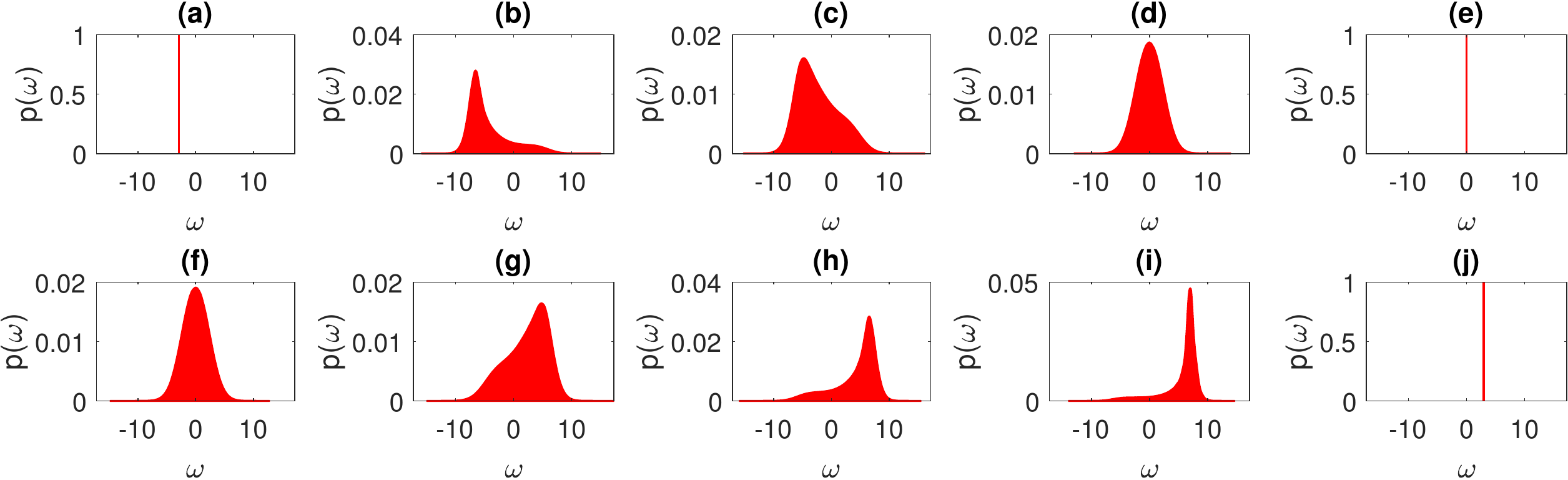}}
\caption{The probability density function of the stationary frequencies of the oscillators in a random network with different frustrations.  Different panels correspond to  (a) $\alpha=0.1\pi$, (b) $\alpha=0.4\pi$, (c) $\alpha=0.45\pi$, (d) $\alpha=0.5\pi$, (e) $\alpha=\pi$, (f) $\alpha=1.5\pi$, (g) $\alpha=1.55\pi$, (h) $\alpha=1.6\pi$, (i) $\alpha=1.64\pi$, and (j) $\alpha=1.9\pi$.  }
\label{DF-RND}
\end{figure}

\begin{figure}[]
   \centering
         \subfigure[\label{F-WS}]{\includegraphics[width=0.49\columnwidth]{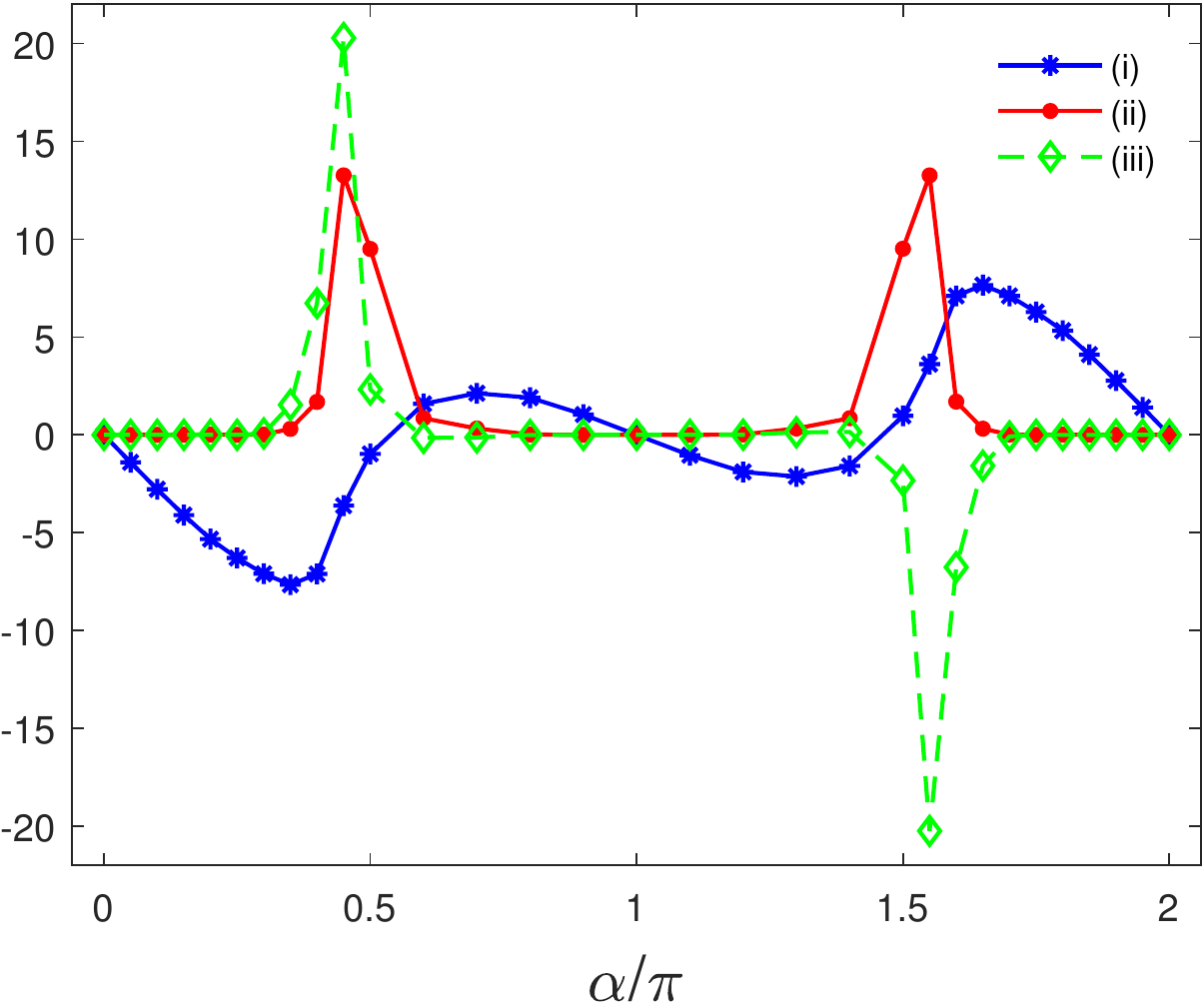}}
         \subfigure[\label{F-R}] {\includegraphics[width=0.49\columnwidth]{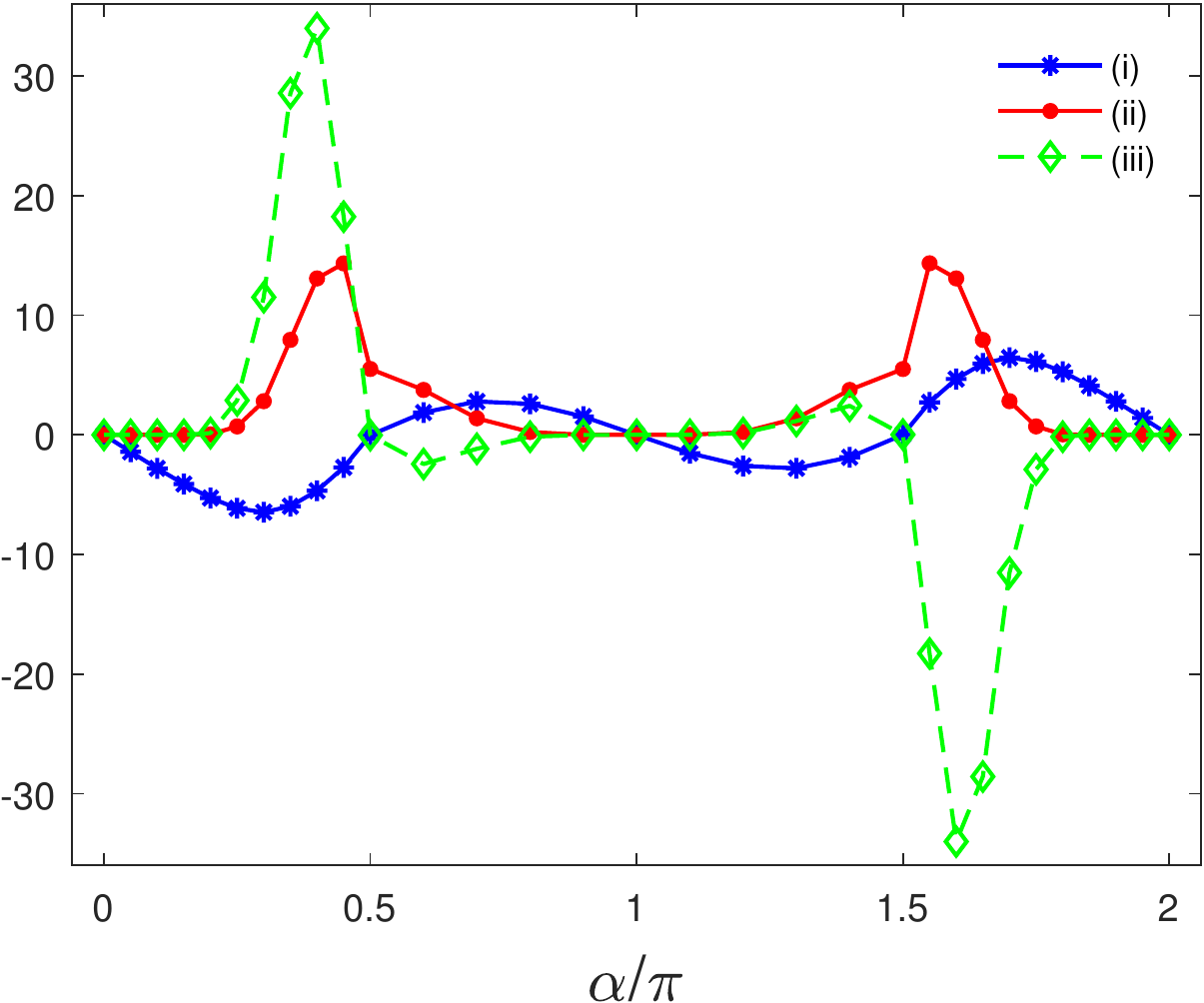}}        
\caption{The long-time average of first~(i), second~(ii) and halved third~(iii) cumulants of the frequency distributions,   as a function of  $\alpha$ for 
(a) SW network with $p=0.03 $, and (b) random network with $p=1$. The first cumulant is the expected value; the second and third cumulants are respectively the variance and  skewness. The simulation parameters in both panels are $N=1000$, $ \langle k \rangle = 10 $.} 
 
\end{figure}

\begin{figure}[]
   \centering
     \subfigure[\label{R-alpha-sw}]{\includegraphics[width=0.49\columnwidth]{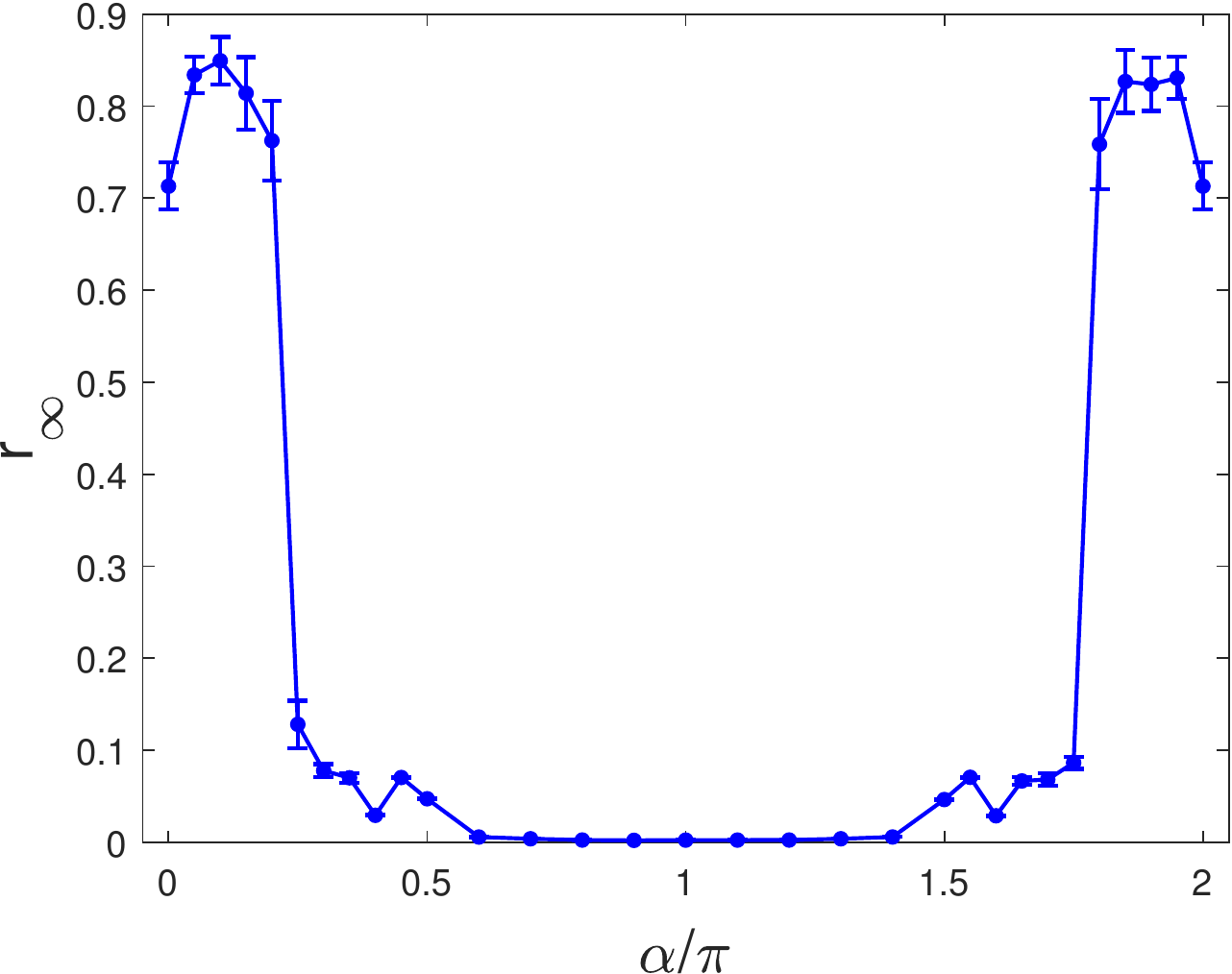}}
     \subfigure[\label{R-alpha-ER}] {\includegraphics[width=0.49\columnwidth]{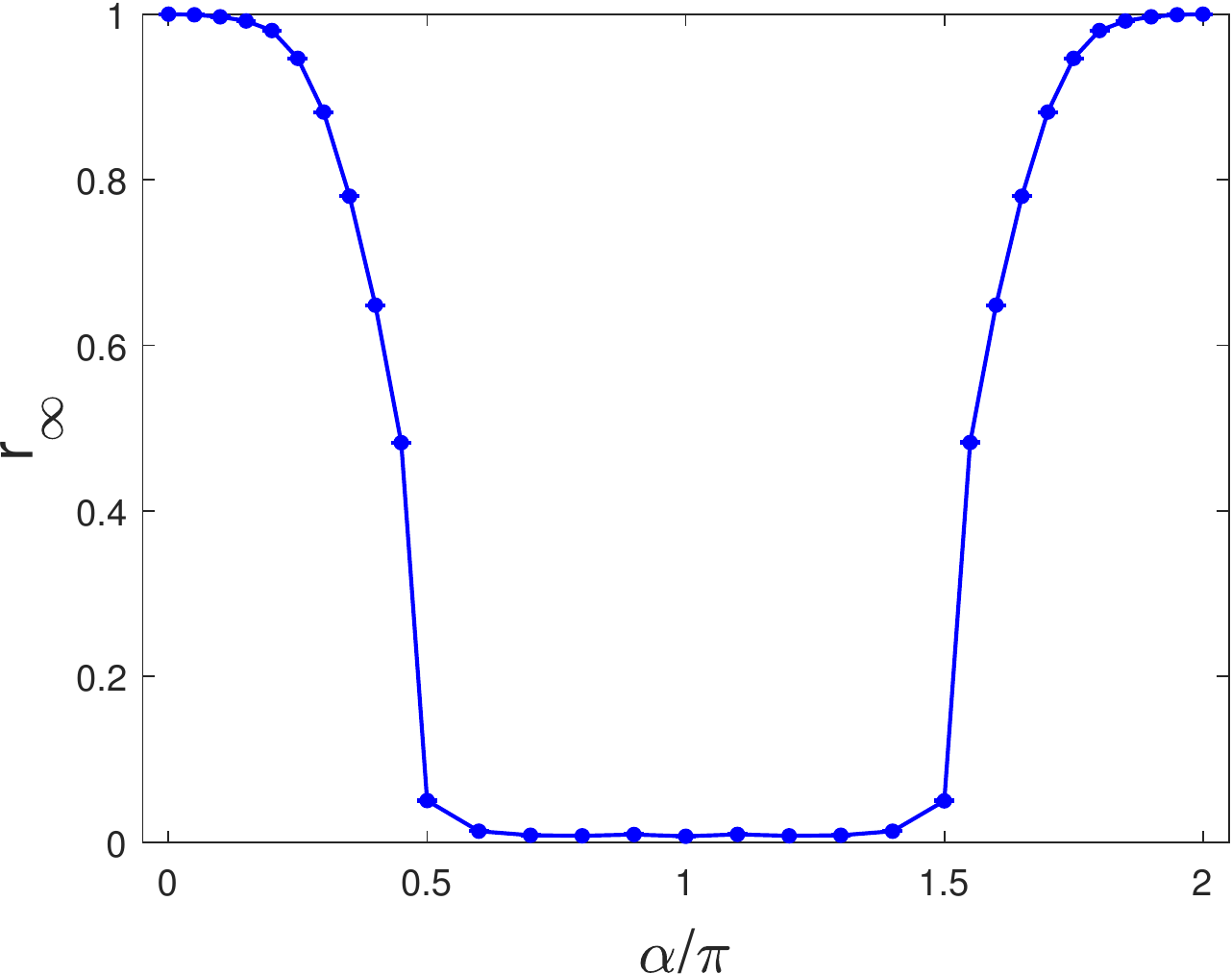}}
\caption{Long-time averaged order parameter ($r_{\infty}$) versus  $\alpha$ for (a)  SW network with $p=0.03$, and (b) random network with $p=1$. The simulation parameters in both panels are $N=1000$, $ \langle k \rangle = 10 $, and $\lambda=1$. The averaging is done over {30 independent initial uniform phase distributions and the error bars denote the standard error of means}.}
\label{R-alpha} 
\end{figure}

\begin{figure}[]
   \centering
         \subfigure[\label{F-B-R-alpha-SW}]{\includegraphics[width=0.49\columnwidth]{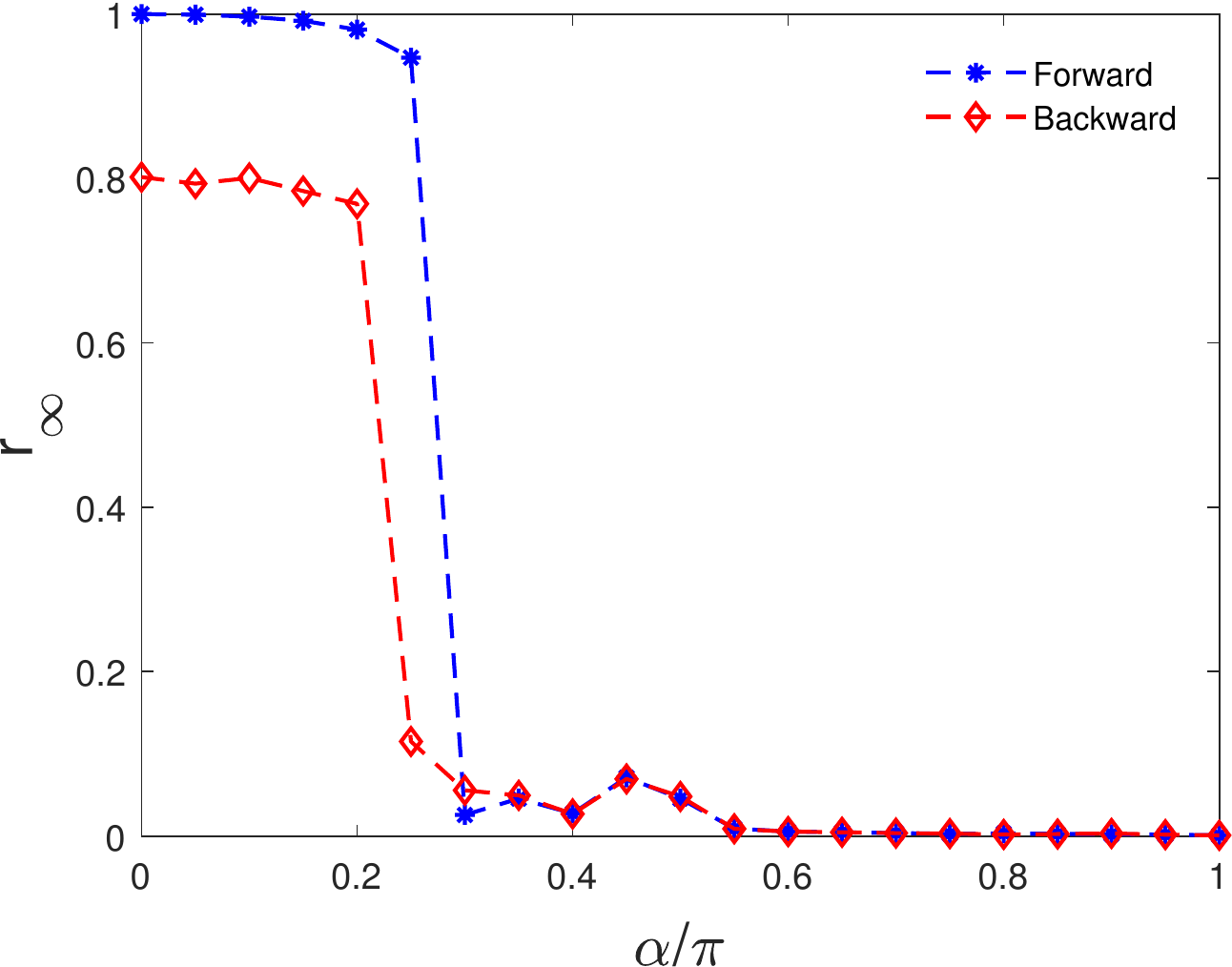}}
         \subfigure[\label{F-B-R-alpha-R}] {\includegraphics[width=0.49\columnwidth]{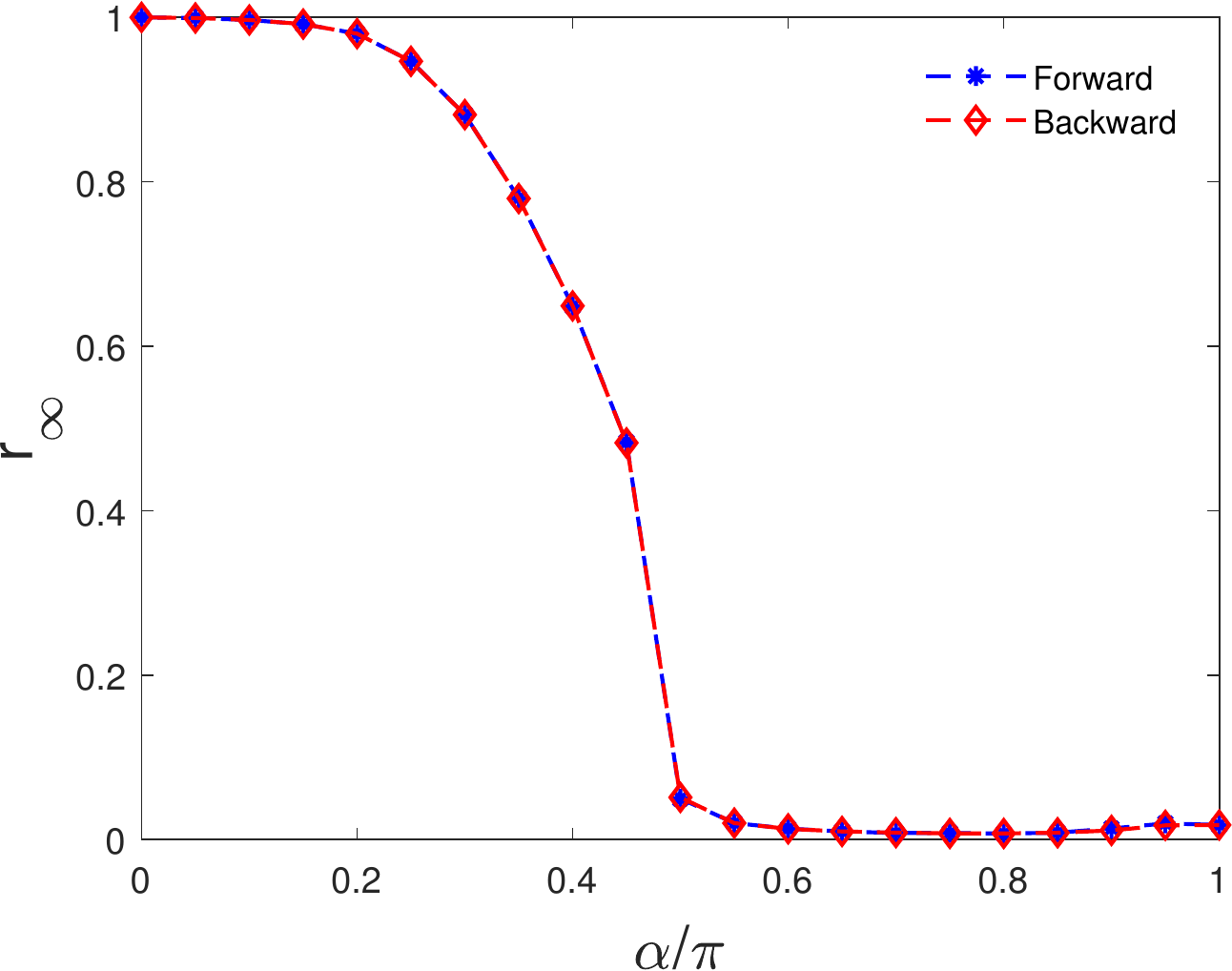}}        
\caption{Forward and backward stationary order parameter ($r_{\infty}$) versus the frustration ($\alpha$) for 
(a)   SW network with $p=0.03 $, and (b) random network with $p=1$. The simulation parameters in both panels are $N=1000$, $ \langle k \rangle =10 $.}
\label{FB} 
\end{figure}

\begin{figure}[]
 \subfigure[\label{chimera}] {\includegraphics[width=0.5\columnwidth]{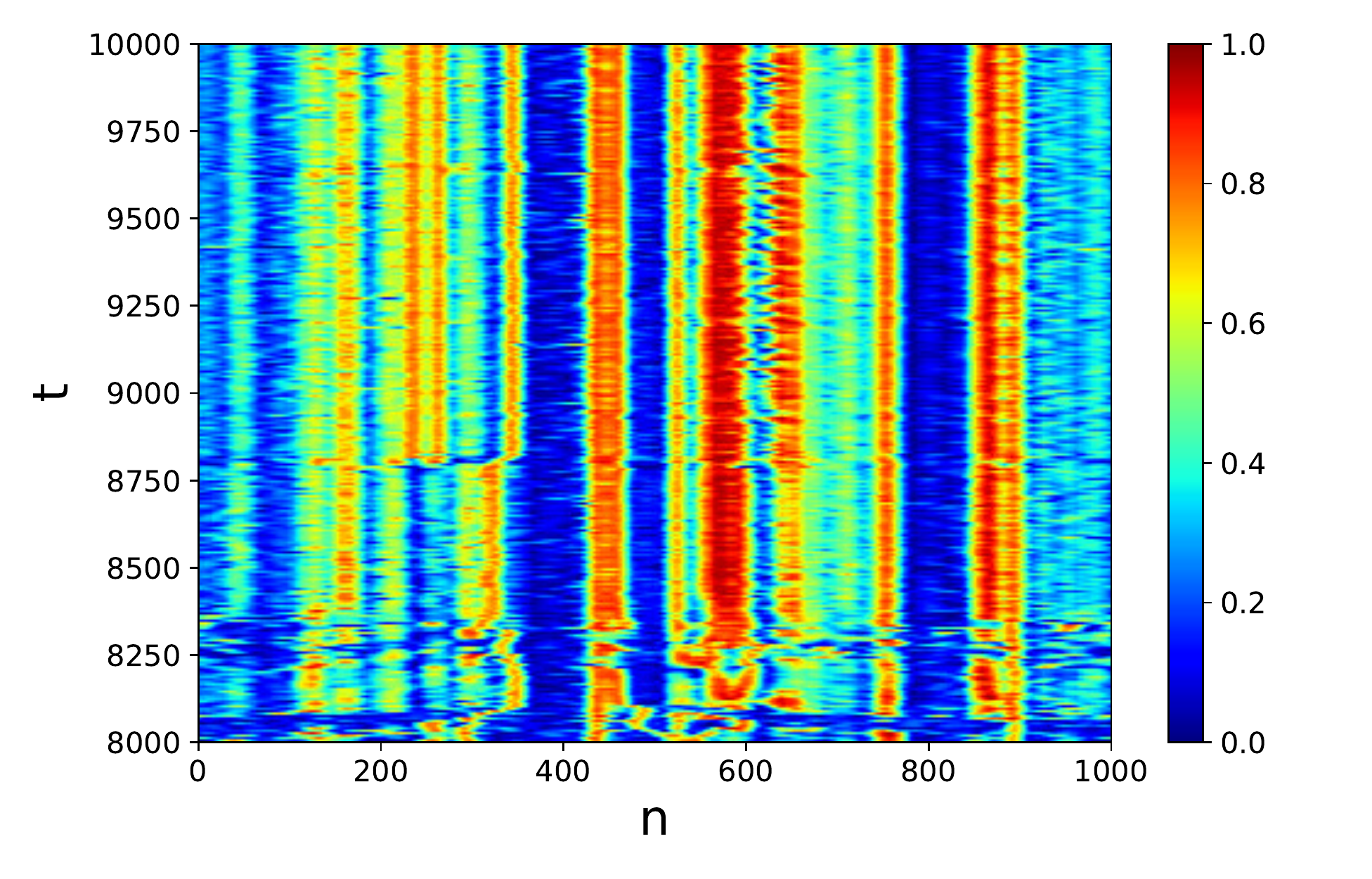}}
 \subfigure[\label{chimerasn}] {\includegraphics[width=0.42\columnwidth]{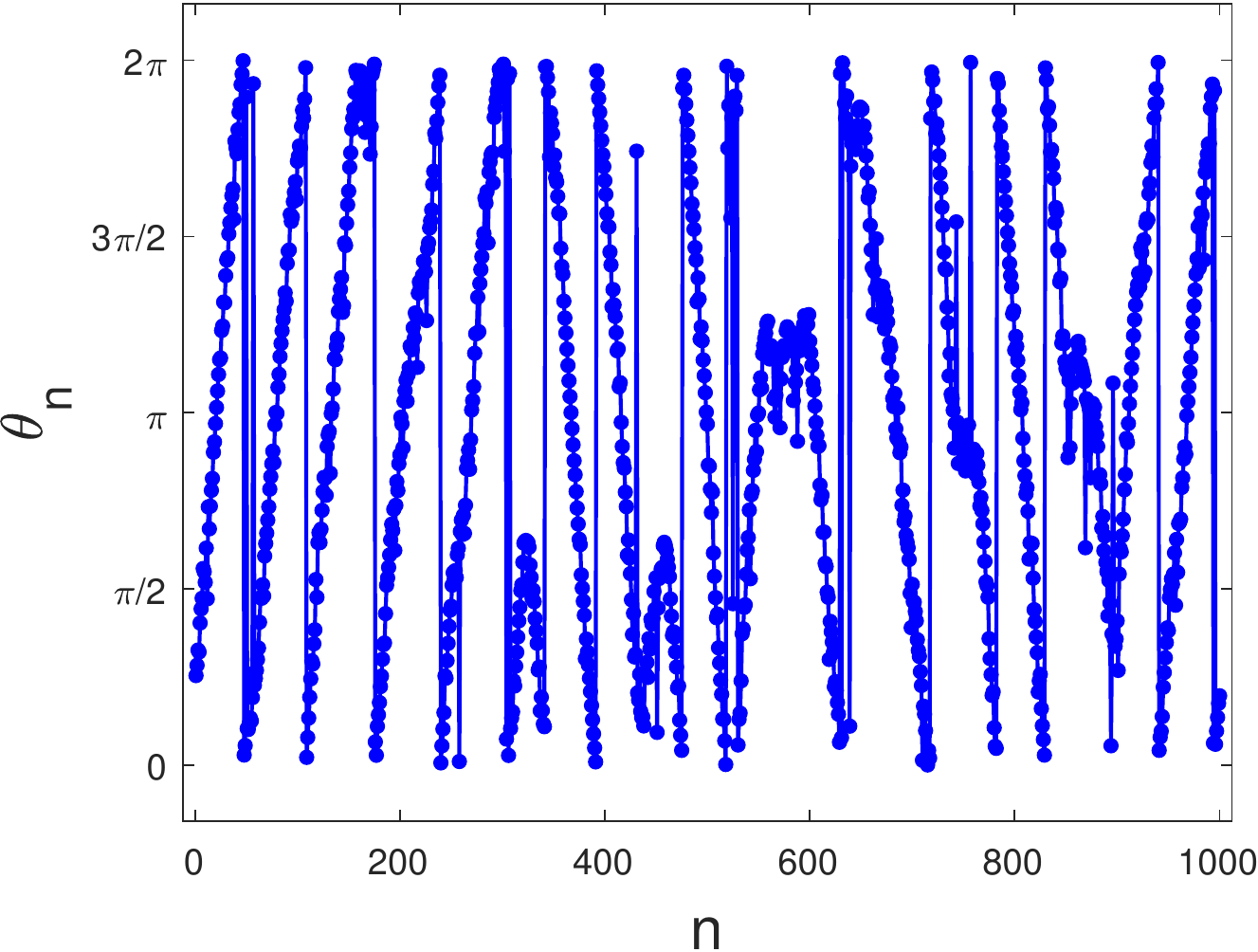}}
\caption{The chimera states in the SW network. (a) Density plot of the local order parameter as defined in Eq.~\eqref{Eq:Rij}. (b) One snapshot of the stationary phase configuration. The simulation parameters are $N=1000$, $ \langle k \rangle = 10 $, $p=0.03$, $\alpha=1.642\pi$,  and $m=20$.}
\end{figure}

Fig.~\ref{r-t-WS} displays the time evolution of the Kuramoto order parameter ($r$) in the SW network for different values of frustration parameter, i.e. $\alpha=0, 0.1\pi, 0.2\pi, 0.3\pi, 0.35\pi, 0.4\pi, 0.5\pi, \pi$. In this figure, one can see that as the frustration parameter increases, the global phase synchrony first rises but then falls, becoming almost zero for large enough $\alpha$ values. As we mentioned before, using linearization error a lower bound has found for the global order parameter of Skaguchi-Kuramoto model which is decreasing by frustration~\cite{ghorban2021linearization}.  These results show that this lower bound is not a good representive of order parameter for the global synchrony of the SW  networks. In other words, in SW networks, a smaller lower bound for the order parameter may correspond to larger synchrony, which is the result of the high linearization error in SW networks.

Fig.~\ref{r-t-R} illustrates a similar plot for the random networks. We can notice that the results obtained for random networks are significantly different. In the random network, the Synchronization is always decreasing as the phase shift $\alpha$ increases from $0$ to $\pi/2$ in. This is in agreement with previous studies, based on linearization of the Sakaguchi-Kuramoto model, indicating that perfect synchronization becomes unattainable in steady-state, even in the limit of infinite coupling strength~\cite{skardal2015erosion}.


To gain more detailed information on the local dynamics, we compute the correlations between all pairs of nodes in the frustrated SW networks using Eq.~\eqref{Eq:dij}. Different panels in Fig.~\ref{CSW} illustrate the heat maps of the time-averaged correlation matrices ($D_{ij}$) for different values of $\alpha$ at the stationary period. One observes various types of collective behavior for different values of frustration parameters. Panel~(a) displays an inhomogeneous phase-locked state with isolated point-like defects for $\alpha=0$. Turning on the frustration makes the steady-state homogeneous by destroying the defects and ending up the system's dynamic to the fully synchronized state for $\alpha=0.1\pi$ (panel~(b)). Further increase of the frustration parameter $\alpha>0.1\pi$ has a destroying effect on the synchronization, so giving rise to decreasing the global synchrony either by stabilizing a quasi-periodic phase-locked state (panels~(d) and (e)) or by reaching the system to a dynamic or static incoherent state (panels~(g) and (h) respectively). Panel~(f) shows the mixture of dynamic and static incoherent states.

However, for the random networks, the heat-maps of the time-averaged correlation matrices at the stationary period shown in Fig.~\ref{CR}, indicate the homogeneous states for all frustration parameters and also  the continuous decline of pair correlations by increasing the frustration and eventually reaching the system to an incoherent state at $\alpha=0.5\pi$.
 

The stationary frequency distribution functions of the oscillators in the SW network, in the rotating frame ($\omega_0$), are plotted  in Fig.~\ref{DF-SW} for different values of $\alpha$. 
In the absence of frustration, all oscillators are phase-locked and rotate with natural frequency ($\omega_0$) in the lab frame. 
A small change in $\alpha$ causes the peak of $P(\omega)$ to shift to the left or right, depending on $\sin(\alpha)$ has a positive or negative value, respectively. An explanation for this comes from the linearization of the frustrated Kuramoto model for the small values of $\alpha$, where the frustrated Kuramoto model turns to the linearized Kuramoto model with frustration-dependent natural frequencies and coupling strength ($\omega'_i=\omega_0-d_i \sin\alpha$). 
Therefore, for small values of $\alpha$ the peak of $P(\omega)$ will be around $\langle \omega'_i\rangle$. The system is in the dynamic incoherent state for $\alpha=\pi/2$ and $\alpha=3\pi/2$, where $P(\omega)$ is a normal distribution with zero mean. The frustrated Kuramoto model with $\alpha=\pi/2$ or $\alpha=3\pi/2$ resembles the phase model with type-1 PRC oscillators. Previous studies have shown that type-1 PRC oscillators do not tend to synchronize via weak mutual excitations~\cite{ermentrout1996type,ziaeemehr2020emergence}. Therefore incoherent collective behavior emerges out of self-organization at $\alpha=\pi/2$ and $\alpha=3\pi/2$. On the other hand, $\alpha=\pi$ corresponds to static incoherent states, where the oscillators are phase-locked at random phases around the unit circle (glassy phase). Indeed, the frustrated Kuramoto model with $\alpha=\pi$ and excitatory interactions corresponds to the ordinary Kuramoto model with inhibitory interactions. That's why the system reaches the static incoherent state  at $\alpha=\pi$. In summary, we can say that frequency distribution is shifted and broadened during the transition from coherent to static random phase-locked states. Qualitatively similar probability density functions of the oscillator's frequency are found for random networks, even though their dynamics are different (see Fig.~\ref{DF-RND}). 

In order to get a better insight into how frustration changes the frequency distribution, mean, variance and skewness of the frequency distribution versus frustration are plotted for SW and random networks~(see Figures \ref{F-WS} and \ref{F-R}). One can see that the average frequency oscillates by varying $\alpha$ in both SW and random networks.

The collective behaviors of oscillators in the two networks are different where $|\Omega|$ reaches its highest point, i.e. $\alpha=0.35\pi$. In fact, oscillators coupled in the random networks are close to synchrony in their fast dynamics. On the other hand, small-world networks enter fast dynamics for the frustration parameter that corresponds to a quasi-periodic state (see panel (e) of the Figs.~\ref{CSW} and \ref{CR}). The curve $\Omega$ versus $\alpha$ is a smooth function for both small-world and random networks, while sharp jumps in the mean frequency  has been reported considering time delay~\cite{ameli2021time2}. We see sharp changes in the variance of the frequency distribution in small-world networks, but not in random networks. In addition, we can find more skewed frequency distributions in random networks than in SW networks. These differences must be related to their structural differences.

 Fig.~\ref{R-alpha} indicates the long-time averaged order parameter in the SW and random networks when $\alpha$ varies from $0$ to $2\pi$. The values of the stationary order parameters were averaged over 30 independent realizations of random initial phases. One can see that the averaged curves are almost symmetric around $\alpha=\pi$. 
However, that is not necessarily true for the curves corresponding to the single realizations. The reason is that, Eq.~\eqref{Eq:sakaguchikuramoto} not being invariant under transformation $(\alpha)\rightarrow (2\pi-\alpha)$. The results show that regardless of the choice of the initial phases, small values of frustration increase the synchrony in the SW but not the random networks. In the interval $\pi/2 \lesssim \alpha \lesssim 3\pi/2$, the order parameter is precisely zero with no fluctuation, indicating the random phase-locked or spin glass states in both SW and random networks.


 Fig.~\ref{FB}-(a) and \ref{FB}-(b) illustrate the variation of the order parameter versus the frustration parameter in forward and backward directions. This figure shows an explosive transition with a hysteresis loop in the dynamics of the SW network, while this is not observed in the random network. This observation suggests the existence of memory in the transition from the coherent state to the incoherent state in the frustrated Kuramoto model in SW networks.

Interestingly, we found the chimera states for some values of the frustrations close to the explosive transition in the SW network. Figs.~\ref{chimera} and \ref{chimerasn} display the evolution of the local order parameter and a phase snapshot at $\alpha=1.64\pi$, where the frequency distribution starts to broaden and skewed. These figures show a few clusters of nearly synchronized phase oscillators, coexisting with incoherent clusters. We need to mention that we did not find such chimera states in the frustrated Kuramoto model in random networks.  
The existence of the Chimera states in an array of identical phase-oscillators with non-local coupling was first discovered by Abrams and Strogatz~\cite{abrams2004chimera}, but it is not well understood yet.  

\section{conclusion}
In summary, we studied the dynamics of identical frustrated Kuramoto oscillators in SW and random networks with a uniform frustration parameter and observed rich dynamical behaviors in the collective dynamics of the SW networks. 
Depending on the value of frustration, the stationary state of the model can be inhomogeneously phase-locked with isolated defects or quasi-periodic patterns,  coherent, partially synchronized, or dynamically incoherent state. Our results show that frustration can derive the dynamics toward synchrony by destroying the defect patterns in SW networks. However, it monotonically decreases synchrony in the random networks.  
The enhancement of synchronization using frustration has
also been reported in previous studies, where the synchrony is achieved by adjusting the frequency and frustration of each node for different network structures~\cite{brede2016frustration}. 
We also found an explosive transition accompanied by a hysteresis loop for the synchronization of SW networks. On the other hand, the phase transition from coherent to incoherent state for the random networks is continuous. We also found the chimera states in the SW networks, for the frustration values close to transition, where the frequency distribution is started to be skewed.  
This work is evidence of the rich dynamical behavior of SW
topologies in which even a population of identical coupled oscillators
 represents novel collective phenomena. We hope our findings lead to a better understanding of real-world networks that are often prone to phase-frustrated couplings. 
\label{conclusion}

\bibliography{bibliography}

\begin{thebibliography}{37}
\expandafter\ifx\csname natexlab\endcsname\relax\def\natexlab#1{#1}\fi
\expandafter\ifx\csname bibnamefont\endcsname\relax
  \def\bibnamefont#1{#1}\fi
\expandafter\ifx\csname bibfnamefont\endcsname\relax
  \def\bibfnamefont#1{#1}\fi
\expandafter\ifx\csname citenamefont\endcsname\relax
  \def\citenamefont#1{#1}\fi
\expandafter\ifx\csname url\endcsname\relax
  \def\url#1{\texttt{#1}}\fi
\expandafter\ifx\csname urlprefix\endcsname\relax\def\urlprefix{URL }\fi
\providecommand{\bibinfo}[2]{#2}
\providecommand{\eprint}[2][]{\url{#2}}

\bibitem[{\citenamefont{Mirollo and
  Strogatz}(1990)}]{mirollo1990synchronization}
\bibinfo{author}{\bibfnamefont{R.~E.} \bibnamefont{Mirollo}} \bibnamefont{and}
  \bibinfo{author}{\bibfnamefont{S.~H.} \bibnamefont{Strogatz}},
  \bibinfo{journal}{SIAM Journal on Applied Mathematics}
  \textbf{\bibinfo{volume}{50}}, \bibinfo{pages}{1645} (\bibinfo{year}{1990}).

\bibitem[{\citenamefont{Garcia-Ojalvo et~al.}(2004)\citenamefont{Garcia-Ojalvo,
  Elowitz, and Strogatz}}]{garcia2004modeling}
\bibinfo{author}{\bibfnamefont{J.}~\bibnamefont{Garcia-Ojalvo}},
  \bibinfo{author}{\bibfnamefont{M.~B.} \bibnamefont{Elowitz}},
  \bibnamefont{and} \bibinfo{author}{\bibfnamefont{S.~H.}
  \bibnamefont{Strogatz}}, \bibinfo{journal}{Proceedings of the National
  Academy of Sciences} \textbf{\bibinfo{volume}{101}}, \bibinfo{pages}{10955}
  (\bibinfo{year}{2004}).

\bibitem[{\citenamefont{Tyson}(1973)}]{tyson1973some}
\bibinfo{author}{\bibfnamefont{J.~J.} \bibnamefont{Tyson}},
  \bibinfo{journal}{The Journal of Chemical Physics}
  \textbf{\bibinfo{volume}{58}}, \bibinfo{pages}{3919} (\bibinfo{year}{1973}).

\bibitem[{\citenamefont{Nijmeijer and
  Rodriguez-Angeles}(2003)}]{nijmeijer2003synchronization}
\bibinfo{author}{\bibfnamefont{H.}~\bibnamefont{Nijmeijer}} \bibnamefont{and}
  \bibinfo{author}{\bibfnamefont{A.}~\bibnamefont{Rodriguez-Angeles}},
  \emph{\bibinfo{title}{Synchronization of mechanical systems}},
  vol.~\bibinfo{volume}{46} (\bibinfo{publisher}{World Scientific},
  \bibinfo{year}{2003}).

\bibitem[{\citenamefont{Blasius et~al.}(1999)\citenamefont{Blasius, Huppert,
  and Stone}}]{blasius1999complex}
\bibinfo{author}{\bibfnamefont{B.}~\bibnamefont{Blasius}},
  \bibinfo{author}{\bibfnamefont{A.}~\bibnamefont{Huppert}}, \bibnamefont{and}
  \bibinfo{author}{\bibfnamefont{L.}~\bibnamefont{Stone}},
  \bibinfo{journal}{Nature} \textbf{\bibinfo{volume}{399}},
  \bibinfo{pages}{354} (\bibinfo{year}{1999}).

\bibitem[{\citenamefont{Nakao}(2016)}]{nakao2016phase}
\bibinfo{author}{\bibfnamefont{H.}~\bibnamefont{Nakao}},
  \bibinfo{journal}{Contemporary Physics} \textbf{\bibinfo{volume}{57}},
  \bibinfo{pages}{188} (\bibinfo{year}{2016}).

\bibitem[{\citenamefont{Kuramoto}(1975)}]{kuramoto1975self}
\bibinfo{author}{\bibfnamefont{Y.}~\bibnamefont{Kuramoto}}, in
  \emph{\bibinfo{booktitle}{International symposium on mathematical problems in
  theoretical physics}} (\bibinfo{organization}{Springer},
  \bibinfo{year}{1975}), pp. \bibinfo{pages}{420--422}.

\bibitem[{\citenamefont{Kuramoto}(2012)}]{kuramoto2012chemical}
\bibinfo{author}{\bibfnamefont{Y.}~\bibnamefont{Kuramoto}},
  \emph{\bibinfo{title}{Chemical oscillations, waves, and turbulence}},
  vol.~\bibinfo{volume}{19} (\bibinfo{publisher}{Springer Science \& Business
  Media}, \bibinfo{year}{2012}).

\bibitem[{\citenamefont{Arenas et~al.}(2008)\citenamefont{Arenas,
  D{\'\i}az-Guilera, Kurths, Moreno, and Zhou}}]{arenas2008synchronization}
\bibinfo{author}{\bibfnamefont{A.}~\bibnamefont{Arenas}},
  \bibinfo{author}{\bibfnamefont{A.}~\bibnamefont{D{\'\i}az-Guilera}},
  \bibinfo{author}{\bibfnamefont{J.}~\bibnamefont{Kurths}},
  \bibinfo{author}{\bibfnamefont{Y.}~\bibnamefont{Moreno}}, \bibnamefont{and}
  \bibinfo{author}{\bibfnamefont{C.}~\bibnamefont{Zhou}},
  \bibinfo{journal}{Physics reports} \textbf{\bibinfo{volume}{469}},
  \bibinfo{pages}{93} (\bibinfo{year}{2008}).

\bibitem[{\citenamefont{Rodrigues et~al.}(2016)\citenamefont{Rodrigues, Peron,
  Ji, and Kurths}}]{rodrigues2016kuramoto}
\bibinfo{author}{\bibfnamefont{F.~A.} \bibnamefont{Rodrigues}},
  \bibinfo{author}{\bibfnamefont{T.~K.~D.} \bibnamefont{Peron}},
  \bibinfo{author}{\bibfnamefont{P.}~\bibnamefont{Ji}}, \bibnamefont{and}
  \bibinfo{author}{\bibfnamefont{J.}~\bibnamefont{Kurths}},
  \bibinfo{journal}{Physics Reports} \textbf{\bibinfo{volume}{610}},
  \bibinfo{pages}{1} (\bibinfo{year}{2016}).

\bibitem[{\citenamefont{Skardal et~al.}(2014)\citenamefont{Skardal, Taylor, and
  Sun}}]{skardal2014optimal}
\bibinfo{author}{\bibfnamefont{P.~S.} \bibnamefont{Skardal}},
  \bibinfo{author}{\bibfnamefont{D.}~\bibnamefont{Taylor}}, \bibnamefont{and}
  \bibinfo{author}{\bibfnamefont{J.}~\bibnamefont{Sun}},
  \bibinfo{journal}{Physical review letters} \textbf{\bibinfo{volume}{113}},
  \bibinfo{pages}{144101} (\bibinfo{year}{2014}).

\bibitem[{\citenamefont{Ghorban et~al.}(2021)\citenamefont{Ghorban, Baharifard,
  Hesaam, Zarei, and Sarbazi-Azad}}]{ghorban2021linearization}
\bibinfo{author}{\bibfnamefont{S.~H.} \bibnamefont{Ghorban}},
  \bibinfo{author}{\bibfnamefont{F.}~\bibnamefont{Baharifard}},
  \bibinfo{author}{\bibfnamefont{B.}~\bibnamefont{Hesaam}},
  \bibinfo{author}{\bibfnamefont{M.}~\bibnamefont{Zarei}}, \bibnamefont{and}
  \bibinfo{author}{\bibfnamefont{H.}~\bibnamefont{Sarbazi-Azad}},
  \bibinfo{journal}{Applied Mathematics and Computation}
  \textbf{\bibinfo{volume}{411}}, \bibinfo{pages}{126464}
  (\bibinfo{year}{2021}).

\bibitem[{\citenamefont{Watts and Strogatz}(1998)}]{watts1998collective}
\bibinfo{author}{\bibfnamefont{D.~J.} \bibnamefont{Watts}} \bibnamefont{and}
  \bibinfo{author}{\bibfnamefont{S.~H.} \bibnamefont{Strogatz}},
  \bibinfo{journal}{nature} \textbf{\bibinfo{volume}{393}},
  \bibinfo{pages}{440} (\bibinfo{year}{1998}).

\bibitem[{\citenamefont{Watts}(2001)}]{watts2001small}
\bibinfo{author}{\bibfnamefont{D.~J.} \bibnamefont{Watts}},
  \emph{\bibinfo{title}{Small worlds: The dynamics of networks between order
  and randomness}} (\bibinfo{publisher}{Princeton University Press},
  \bibinfo{year}{2001}).

\bibitem[{\citenamefont{Barahona and
  Pecora}(2002)}]{barahona2002synchronization}
\bibinfo{author}{\bibfnamefont{M.}~\bibnamefont{Barahona}} \bibnamefont{and}
  \bibinfo{author}{\bibfnamefont{L.~M.} \bibnamefont{Pecora}},
  \bibinfo{journal}{Physical review letters} \textbf{\bibinfo{volume}{89}},
  \bibinfo{pages}{054101} (\bibinfo{year}{2002}).

\bibitem[{\citenamefont{Esfahani et~al.}(2012)\citenamefont{Esfahani, Shahbazi,
  and Samani}}]{esfahanii2012noise}
\bibinfo{author}{\bibfnamefont{R.~K.} \bibnamefont{Esfahani}},
  \bibinfo{author}{\bibfnamefont{F.}~\bibnamefont{Shahbazi}}, \bibnamefont{and}
  \bibinfo{author}{\bibfnamefont{K.~A.} \bibnamefont{Samani}},
  \bibinfo{journal}{Physical Review E} \textbf{\bibinfo{volume}{86}},
  \bibinfo{pages}{036204} (\bibinfo{year}{2012}).

\bibitem[{\citenamefont{Nikfard et~al.}(2021)\citenamefont{Nikfard, Tabatabaei,
  Esfahani, and Shahbazi}}]{nikfard2021enhancement}
\bibinfo{author}{\bibfnamefont{T.}~\bibnamefont{Nikfard}},
  \bibinfo{author}{\bibfnamefont{Y.~H.} \bibnamefont{Tabatabaei}},
  \bibinfo{author}{\bibfnamefont{R.~K.} \bibnamefont{Esfahani}},
  \bibnamefont{and} \bibinfo{author}{\bibfnamefont{F.}~\bibnamefont{Shahbazi}},
  \bibinfo{journal}{The European Physical Journal Plus}
  \textbf{\bibinfo{volume}{136}}, \bibinfo{pages}{1} (\bibinfo{year}{2021}).

\bibitem[{\citenamefont{Ameli et~al.}(2021)\citenamefont{Ameli, Karimian, and
  Shahbazi}}]{ameli2021time2}
\bibinfo{author}{\bibfnamefont{S.}~\bibnamefont{Ameli}},
  \bibinfo{author}{\bibfnamefont{M.}~\bibnamefont{Karimian}}, \bibnamefont{and}
  \bibinfo{author}{\bibfnamefont{F.}~\bibnamefont{Shahbazi}},
  \bibinfo{journal}{Chaos: An Interdisciplinary Journal of Nonlinear Science}
  \textbf{\bibinfo{volume}{31}}, \bibinfo{pages}{113125}
  (\bibinfo{year}{2021}).

\bibitem[{\citenamefont{Yokoi et~al.}(1988)\citenamefont{Yokoi, Tang, and
  Chou}}]{yokoi1988ground}
\bibinfo{author}{\bibfnamefont{C.~S.} \bibnamefont{Yokoi}},
  \bibinfo{author}{\bibfnamefont{L.-H.} \bibnamefont{Tang}}, \bibnamefont{and}
  \bibinfo{author}{\bibfnamefont{W.}~\bibnamefont{Chou}},
  \bibinfo{journal}{Physical Review B} \textbf{\bibinfo{volume}{37}},
  \bibinfo{pages}{2173} (\bibinfo{year}{1988}).

\bibitem[{\citenamefont{Zheng et~al.}(1998)\citenamefont{Zheng, Hu, and
  Hu}}]{zheng1998resonant}
\bibinfo{author}{\bibfnamefont{Z.}~\bibnamefont{Zheng}},
  \bibinfo{author}{\bibfnamefont{B.}~\bibnamefont{Hu}}, \bibnamefont{and}
  \bibinfo{author}{\bibfnamefont{G.}~\bibnamefont{Hu}},
  \bibinfo{journal}{Physical Review B} \textbf{\bibinfo{volume}{58}},
  \bibinfo{pages}{5453} (\bibinfo{year}{1998}).

\bibitem[{\citenamefont{Watanabe et~al.}(1996)\citenamefont{Watanabe, van~der
  Zant, Strogatz, and Orlando}}]{watanabe1996dynamics}
\bibinfo{author}{\bibfnamefont{S.}~\bibnamefont{Watanabe}},
  \bibinfo{author}{\bibfnamefont{H.~S.} \bibnamefont{van~der Zant}},
  \bibinfo{author}{\bibfnamefont{S.~H.} \bibnamefont{Strogatz}},
  \bibnamefont{and} \bibinfo{author}{\bibfnamefont{T.~P.}
  \bibnamefont{Orlando}}, \bibinfo{journal}{Physica D: Nonlinear Phenomena}
  \textbf{\bibinfo{volume}{97}}, \bibinfo{pages}{429} (\bibinfo{year}{1996}).

\bibitem[{\citenamefont{Sakaguchi et~al.}(1988)\citenamefont{Sakaguchi,
  Shinomoto, and Kuramoto}}]{sakaguchi1988mutual}
\bibinfo{author}{\bibfnamefont{H.}~\bibnamefont{Sakaguchi}},
  \bibinfo{author}{\bibfnamefont{S.}~\bibnamefont{Shinomoto}},
  \bibnamefont{and} \bibinfo{author}{\bibfnamefont{Y.}~\bibnamefont{Kuramoto}},
  \bibinfo{journal}{Progress of theoretical physics}
  \textbf{\bibinfo{volume}{79}}, \bibinfo{pages}{1069} (\bibinfo{year}{1988}).

\bibitem[{\citenamefont{Sakaguchi and Kuramoto}(1986)}]{sakaguchi1986soluble}
\bibinfo{author}{\bibfnamefont{H.}~\bibnamefont{Sakaguchi}} \bibnamefont{and}
  \bibinfo{author}{\bibfnamefont{Y.}~\bibnamefont{Kuramoto}},
  \bibinfo{journal}{Progress of Theoretical Physics}
  \textbf{\bibinfo{volume}{76}}, \bibinfo{pages}{576} (\bibinfo{year}{1986}).

\bibitem[{\citenamefont{Ha et~al.}(2018)\citenamefont{Ha, Kim, and
  Park}}]{ha2018remarks}
\bibinfo{author}{\bibfnamefont{S.-Y.} \bibnamefont{Ha}},
  \bibinfo{author}{\bibfnamefont{H.~K.} \bibnamefont{Kim}}, \bibnamefont{and}
  \bibinfo{author}{\bibfnamefont{J.}~\bibnamefont{Park}},
  \bibinfo{journal}{Analysis and Applications} \textbf{\bibinfo{volume}{16}},
  \bibinfo{pages}{525} (\bibinfo{year}{2018}).

\bibitem[{\citenamefont{Crook et~al.}(1997)\citenamefont{Crook, Ermentrout,
  Vanier, and Bower}}]{crook1997role}
\bibinfo{author}{\bibfnamefont{S.~M.} \bibnamefont{Crook}},
  \bibinfo{author}{\bibfnamefont{G.~B.} \bibnamefont{Ermentrout}},
  \bibinfo{author}{\bibfnamefont{M.~C.} \bibnamefont{Vanier}},
  \bibnamefont{and} \bibinfo{author}{\bibfnamefont{J.~M.} \bibnamefont{Bower}},
  \bibinfo{journal}{Journal of computational neuroscience}
  \textbf{\bibinfo{volume}{4}}, \bibinfo{pages}{161} (\bibinfo{year}{1997}).

\bibitem[{\citenamefont{Skardal et~al.}(2015)\citenamefont{Skardal, Taylor,
  Sun, and Arenas}}]{skardal2015erosion}
\bibinfo{author}{\bibfnamefont{P.~S.} \bibnamefont{Skardal}},
  \bibinfo{author}{\bibfnamefont{D.}~\bibnamefont{Taylor}},
  \bibinfo{author}{\bibfnamefont{J.}~\bibnamefont{Sun}}, \bibnamefont{and}
  \bibinfo{author}{\bibfnamefont{A.}~\bibnamefont{Arenas}},
  \bibinfo{journal}{Physical Review E} \textbf{\bibinfo{volume}{91}},
  \bibinfo{pages}{010802} (\bibinfo{year}{2015}).

\bibitem[{\citenamefont{Omel’chenko and
  Wolfrum}(2012)}]{omel2012nonuniversal}
\bibinfo{author}{\bibfnamefont{E.}~\bibnamefont{Omel’chenko}}
  \bibnamefont{and} \bibinfo{author}{\bibfnamefont{M.}~\bibnamefont{Wolfrum}},
  \bibinfo{journal}{Physical review letters} \textbf{\bibinfo{volume}{109}},
  \bibinfo{pages}{164101} (\bibinfo{year}{2012}).

\bibitem[{\citenamefont{Brede and Kalloniatis}(2016)}]{brede2016frustration}
\bibinfo{author}{\bibfnamefont{M.}~\bibnamefont{Brede}} \bibnamefont{and}
  \bibinfo{author}{\bibfnamefont{A.~C.} \bibnamefont{Kalloniatis}},
  \bibinfo{journal}{Physical Review E} \textbf{\bibinfo{volume}{93}},
  \bibinfo{pages}{062315} (\bibinfo{year}{2016}).

\bibitem[{\citenamefont{Sch{\"o}ll}(2016)}]{scholl2016synchronization}
\bibinfo{author}{\bibfnamefont{E.}~\bibnamefont{Sch{\"o}ll}},
  \bibinfo{journal}{The European Physical Journal Special Topics}
  \textbf{\bibinfo{volume}{225}}, \bibinfo{pages}{891} (\bibinfo{year}{2016}).

\bibitem[{\citenamefont{Abrams and Strogatz}(2004)}]{abrams2004chimera}
\bibinfo{author}{\bibfnamefont{D.~M.} \bibnamefont{Abrams}} \bibnamefont{and}
  \bibinfo{author}{\bibfnamefont{S.~H.} \bibnamefont{Strogatz}},
  \bibinfo{journal}{Physical review letters} \textbf{\bibinfo{volume}{93}},
  \bibinfo{pages}{174102} (\bibinfo{year}{2004}).

\bibitem[{\citenamefont{Abrams et~al.}(2008)\citenamefont{Abrams, Mirollo,
  Strogatz, and Wiley}}]{abrams2008solvable}
\bibinfo{author}{\bibfnamefont{D.~M.} \bibnamefont{Abrams}},
  \bibinfo{author}{\bibfnamefont{R.}~\bibnamefont{Mirollo}},
  \bibinfo{author}{\bibfnamefont{S.~H.} \bibnamefont{Strogatz}},
  \bibnamefont{and} \bibinfo{author}{\bibfnamefont{D.~A.} \bibnamefont{Wiley}},
  \bibinfo{journal}{Physical review letters} \textbf{\bibinfo{volume}{101}},
  \bibinfo{pages}{084103} (\bibinfo{year}{2008}).

\bibitem[{\citenamefont{Laing}(2009)}]{laing2009chimera}
\bibinfo{author}{\bibfnamefont{C.~R.} \bibnamefont{Laing}},
  \bibinfo{journal}{Chaos: An Interdisciplinary Journal of Nonlinear Science}
  \textbf{\bibinfo{volume}{19}}, \bibinfo{pages}{013113}
  (\bibinfo{year}{2009}).

\bibitem[{\citenamefont{Yeldesbay et~al.}(2014)\citenamefont{Yeldesbay,
  Pikovsky, and Rosenblum}}]{yeldesbay2014chimeralike}
\bibinfo{author}{\bibfnamefont{A.}~\bibnamefont{Yeldesbay}},
  \bibinfo{author}{\bibfnamefont{A.}~\bibnamefont{Pikovsky}}, \bibnamefont{and}
  \bibinfo{author}{\bibfnamefont{M.}~\bibnamefont{Rosenblum}},
  \bibinfo{journal}{Physical review letters} \textbf{\bibinfo{volume}{112}},
  \bibinfo{pages}{144103} (\bibinfo{year}{2014}).

\bibitem[{\citenamefont{Martens et~al.}(2016)\citenamefont{Martens, Bick, and
  Panaggio}}]{martens2016chimera}
\bibinfo{author}{\bibfnamefont{E.~A.} \bibnamefont{Martens}},
  \bibinfo{author}{\bibfnamefont{C.}~\bibnamefont{Bick}}, \bibnamefont{and}
  \bibinfo{author}{\bibfnamefont{M.~J.} \bibnamefont{Panaggio}},
  \bibinfo{journal}{Chaos: An Interdisciplinary Journal of Nonlinear Science}
  \textbf{\bibinfo{volume}{26}}, \bibinfo{pages}{094819}
  (\bibinfo{year}{2016}).

\bibitem[{\citenamefont{G{\'o}mez-Gardenes
  et~al.}(2007)\citenamefont{G{\'o}mez-Gardenes, Moreno, and
  Arenas}}]{gomez2007paths}
\bibinfo{author}{\bibfnamefont{J.}~\bibnamefont{G{\'o}mez-Gardenes}},
  \bibinfo{author}{\bibfnamefont{Y.}~\bibnamefont{Moreno}}, \bibnamefont{and}
  \bibinfo{author}{\bibfnamefont{A.}~\bibnamefont{Arenas}},
  \bibinfo{journal}{Physical review letters} \textbf{\bibinfo{volume}{98}},
  \bibinfo{pages}{034101} (\bibinfo{year}{2007}).

\bibitem[{\citenamefont{Ermentrout}(1996)}]{ermentrout1996type}
\bibinfo{author}{\bibfnamefont{B.}~\bibnamefont{Ermentrout}},
  \bibinfo{journal}{Neural computation} \textbf{\bibinfo{volume}{8}},
  \bibinfo{pages}{979} (\bibinfo{year}{1996}).

\bibitem[{\citenamefont{Ziaeemehr et~al.}(2020)\citenamefont{Ziaeemehr, Zarei,
  and Sheshbolouki}}]{ziaeemehr2020emergence}
\bibinfo{author}{\bibfnamefont{A.}~\bibnamefont{Ziaeemehr}},
  \bibinfo{author}{\bibfnamefont{M.}~\bibnamefont{Zarei}}, \bibnamefont{and}
  \bibinfo{author}{\bibfnamefont{A.}~\bibnamefont{Sheshbolouki}},
  \bibinfo{journal}{Scientific reports} \textbf{\bibinfo{volume}{10}},
  \bibinfo{pages}{1} (\bibinfo{year}{2020}).

\end{thebibliography}

\end{document}